\documentclass[12pt]{article}

\usepackage[letterpaper, total={6.5in, 9in}]{geometry}
\usepackage{multirow}
\usepackage{rotating}

\newcommand{\ds}{\displaystyle}
\newcommand{\Eq}[1]{Eq.~(\ref{#1})}
\newcommand{\Fig}[1]{Fig.~\ref{#1}}

\newcommand{\code}[1]{{\small\tt #1}}

\begin{document}

\title{Exact Calculation of Expected Values for Splitting Pairs in Blackjack }

\markboth{Splitting Pairs in Blackjack}{Splitting Pairs in Blackjack}

\author{John A. Nairn \\ \small\emph{Oregon State University, Wood Science \&\ Engineering,
112 Richardson Hall}, \\ \small\emph{Corvallis, OR 97330, USA}}

\maketitle

\begin{abstract}
Computer calculations for most exact expected values in blackjack have been available since the 1960's, but exact results for pair splitting and resplitting have previously been too computer intensive. This paper describes a new algorithm for exact pair-splitting. By using dealer probability caching methods and revising the method for recursively generating possible player hands, the estimated calculation time compared to standard methods was reduced by five orders of magnitude. The resulting algorithm was used to calculate the first exact and complete pair splitting results for a single deck game. The exact results were compared to prior approximate theories for resplitting. The prior theories are accurate for many calculations, but inaccurate for resplitting tens. A new approximation method was developed that is accurate for all resplitting calculations.

%\begin{keywords}blackjack; pair splitting; recursive algorithms
%\end{keywords}

\end{abstract}

\section{Introduction}
\label{Introduction}

Many books and articles describe calculation of expected values in blackjack by computer methods. The first accurate derivation of correct strategy was done by Baldwin {\em et al.\/}~\cite{Baldwin:1956}. Due to limited computer capabilities at that time, they used several approximations. In the 1960's, it was realized the problem was sufficiently small that all hands could be generated and therefore exact expected values could be calculated. Tables of results are given in Thorp \cite{Thorp:1962}\ and Epstein \cite{Epstein:1977}.

Most blackjack calculations can be completed in seconds. The one exception is for pair splitting. When a player's first two cards match, that player is permitted to split the cards and play two separate hands. If another matching card is dealt to one of the separate hands, some casinos permit resplitting to make a third or fourth, or potentially even more hands in multideck games. For $N$ possible player hands, computer generation of up to $h$ split hands requires on the order of $N^h$ hands. This calculation is feasible for $h=2$ but because $N$ is large, it becomes computer intensive for $h\ge3$.

There are few published results of pair splitting expected values and none of them are exact. This paper considers the computer problem for finding exact expected values for pair splitting in a single-deck game for $h=2$, 3, and 4. Initial calculations using conventional blackjack computational methods led to an estimate that serial calculations for $h=4$ would require about $3.4\times 10^{11}$ seconds or 11,000 years of CPU time. By using memory caching and revising the approach to enumerating split hands, total calculation time was reduced by five orders of magnitude to $3.9\times 10^{6}$ seconds or 45 days. This paper describes an algorithm that was optimization based on performance in serial calculations. A simple approach to parallel calculations is outlined.

The derived algorithm was used to calculate a table of exact splitting expected values for all rule options. The exact results were compared to a prior approximate approach developed by Griffin \cite{Griffin:1979}, and to other tables of splitting results. Griffin's methods are extremely accurate when resplitting is not allowed, but have some deficiencies when resplitting is allowed. A new approximate method was developed that corrects those deficiencies and agrees with nearly all exact results with absolute errors less than $\pm 0.001$. All other published splitting results deviate more from exact results

\section{Exact Splitting Algorithm}

\subsection{Blackjack and Splitting Rules}

This section briefly describes casino blackjack rules with emphasis on defining terms and on various options for pair splitting. The reader is referred to many references for more details ({\em e.g.}, \cite{Thorp:1962,Thorp:1961}). A hand starts by a player making a wager --- the initial bet size. Cards are then dealt from a single deck or from multiple decks. The player and dealer each receive two cards. One of the dealer's cards is exposed and therefore known to the player (the dealer ``up card''). Cards 2 through 9 are valued with their number; 10's and all face cards are valued 10; aces may be valued 1 or 11 at the player's option. A hand's score is a sum of values of its cards. A hand with no aces is a ``hard'' hand. A hand with an ace that does not exceed 21 when the ace is counted as 11 is a ``soft'' hand ({\em e.g.}, (A,8) is a soft 19). If all aces in a hand must be counted as 1 to avoid exceeding 21, it becomes a hard hand ({\em e.g.}, (8,7,A) is a hard 16).

If the dealer's up card is an ace or a ten card, the ``hole card'' (\emph{i.e.}, the hidden card) is checked to see if the dealer has a ``natural'' or blackjack defined as an ace and a ten card adding to 21. If the dealer has a natural, the hand is over. The player loses the initial bet, unless the player also has a natural, the hand is a draw or ``push'' with no loss or gain to the player. If the dealer does not have a natural, play continues. A player with a natural, wins 1.5 times the initial bet. A player that does not have a natural, can ``stand,'' ``hit,'' or ``double down''. When a player stands, that player is done. A player that hits can continue to hit while the hand total is less than or equal to 21 or stop play at any time by standing. If the final total exceeds 21, the hand is a ``bust'' and the initial bet is lost regardless of the dealer's final hand. A player that doubles down, doubles the initial bet size and then receives one and only one additional card. Some casinos allow the player to double down on any two cards; others restrict doubling down to only hands of hard 10 or 11.

If a player finishes without busting, the dealer plays out the hand by fixed rules regardless of the player's hand. The dealer exposes the hole card and then takes additional cards until the score is 17 or greater. In some casinos, the dealer will hit a soft 17 but stand on all other hands 17 or greater. When done, the dealer's score is compared to the player's. The player wins the current bet size if the dealer busts or if the player's score is higher than the dealer's. The player loses the initial bet for scores lower than the dealer's score. If the scores are equal, the hand is a push.

This paper is about an additional player option known as pair splitting. If the first two cards are matched ({\em e.g.}, hands with (8,8), (4,4), or (10,Q); the last example indicates that any two ten cards can be split), the player can split the hand. When splitting, the two cards are separated into two hands and a second bet equal to the initial bet size in placed on the second hand. These hands get a second card and play continues as for non-split hands with the following exceptions. If first two cards of a split hand total 21, the hand is not a natural, but just a hand totaling 21. When splitting aces, the player receives one additional card on each hand, but cannot continue with other options. At the two-card stage, the rules between casinos vary. Some casinos do not allow doubling down on such hands. Other casinos allow doubling down on any two-card hand or perhaps only on hands of hard 10 or 11. The rules also vary if the second card matches the original split card. Some casinos allow resplitting, while others do not. For those that allow resplitting, that option is often limited to a maximum of 4 hands or occasionally is allowed for any number of hands (splitting to more than four hands is possible when splitting ten cards or in multideck games). Normally aces cannot be resplit even in casinos that allow resplitting. In rare cases, aces can be resplit and the additional hands again receive just one card.

A complete pair-splitting analysis requires analysis of all double down options:
\begin{enumerate}
\item ND --- No double down after splitting
\item DD1 --- Double down on any two cards after splitting
\item DD2 --- Double down only on hard 10 or 11 after splitting
\end{enumerate}
To analyze resplitting options as well, the splitting process is allowed to proceed to a maximum of $h$ hands. Setting $h=2$ gives results when resplitting is not allowed. When $h>2$, resplitting is allowed. Here the goal was to complete calculations for $h=4$, which corresponds to a common limitation in casinos that allow resplitting. In a single deck game, $h=4$ is also unlimited splitting except when splitting 10's.

\subsection{Exact Expected Values}
\label{ExactExpectedValues}

The expected value for any strategy decision is defined as the expected win per unit bet for making a decision and then completing the hand by the zero-memory, basic strategy. Zero-memory basic strategy is defined as the decision that maximizes expected value based only on knowledge of the dealer's up card and the player's initial two cards \cite{Epstein:1977}. The zero-memory basic strategy for any number of decks and for dealers that stand or hit soft 17 is given in Griffin \cite{Griffin:1979}; for completeness, the single-desk basic strategy used for these calculations is in Appendix A.

The exact expected value when choosing to hit a hand, $\vec h$, against dealer up card $u$, can be expressed in a recursive equation as:
\begin{equation}
    E(\vec h,u) = \sum_{i=1}^{10} p(i) \left\{ \begin{array}{ll}
              E(\vec h+i,u) & {\rm if}\ \vec h+i\ {\rm should\ be\ hit} \\
              S(\vec h+i,u) & {\rm if}\ \vec h+i\ {\rm should\ stand}
            \end{array} \right.         \label{exactexval}
\end{equation}
Here $\vec h=(c_1, c_2, c_3,\dots)$ is a player hand with cards $c_1$, $c_2$, \emph{etc.}, $\vec h+i$ is the new hand formed by adding card $i$ to $\vec h$, $p(i)$ is the probability that the next card in the deck has value $i$, $S(\vec h+i,u)$ is the expected value if the player stands on the new hand, and ``should be hit'' or ``should stand'' refers to basic strategy decision based on the new hand $\vec h+i$. A recursive calculation terminates when hand $\vec h+i$ should stand for all values of $i$ because the exact expected value when choosing to stand is given by a closed-form, non-recursive equation:
\[
      S(\vec h,u) = \left\{ \begin{array}{ll}
             -1 & {\rm if}\ s(\vec h)>21 \\
             d(>21)+ \sum_{i=17}^{i<s(\vec h)} d(i) - \sum_{i=s(\vec h)+1}^{i<=21} d(i) & {\rm otherwise}
          \end{array} \right.
\]
Here $s(\vec h)$ is the score for hand $\vec h$, $d(i)$ for $i=17$ to $21$ is the probability the final dealer hand totals $i$, and $d(>21)$ is the probability the dealer busts with a score greater than 21. The first option means the player loses one bet size for a busted hand. Otherwise the player wins one unit if the dealer busts or if $s(\vec h)$ is higher than the dealer's score, loses one if the dealer's score is higher (without busting), or ties if the dealer's and player's scores are equal.

Because expected values during hand play imply the dealer does not have a natural, the player card probabilities, $p(i)$, are conditional on this fact. The conditional $p(i)$ from Griffin \cite{Griffin:1979}\ are:
\begin{equation}\renewcommand{\arraystretch}{1.5}
    p(i) = \left\{ \begin{array}{ll}
        {\ds n_i\over \ds n_{deck}} & {\rm for}\ 2\le u \le 9 \\
        {\ds n_i\over \ds n_{deck}-1} & {\rm for}\ u=1,10\ {\rm and}\ i=11-u  \\
       {\ds n_i\over \ds n_{deck}-1}\left({\ds n_{deck}-n_{11-u}-1\over \ds n_{deck}-n_{11-u}}\right)
                & {\rm for}\ u=1,10\ {\rm and}\ i\ne11-u
         \end{array} \right.         \label{playerpi}
\end{equation}
Here $n_i$ is the number cards remaining in the deck with value $i$ and $n_{deck}$ is the total number of cards remaining in the deck.

Dealer probabilities required for $S(\vec h,u)$ can be expressed as a vector function
\[
    \vec D(u, \vec r) = \bigl( d(17), d(18), d(19), d(20), d(21), d(>21) \bigr)
\]
This function depends on the dealer up card and on the set of cards (besides $u$), $\vec r$, that have been removed from the deck. The probabilities can be calculated by a simple recursive algorithm \cite{Griffin:1979}. Like player card probabilities, dealer probabilities are always made conditional on the dealer not having a natural.

\subsection{Recursive Hand Generation for Pair Splitting}
\label{RecursiveHandGenerationforPairSplitting}

Exact expected value calculations require consideration of all possible hands and each hand (or set of hands when splitting) requires a new calculation of dealer probabilities, because they depend on the contents of the player's hands. A common approach is to use recursive subroutines to generate all hands and sum the results for the exact expected values. This section outlines C++ code that extends this approach to recursive generation of all possible split hands to any maximum number of hands (a software project with complete source code is available on GitHub \cite{Nairn:BJGit}). The code is based on three objects called \code{Deck}, \code{Dealer}, and \code{Hand}. The  \code{Deck} object is responsible for tracking cards in the deck and calculating player weights, $p(i)$. The \code{Dealer} object is responsible for calculating dealer probabilities, $\vec D(u, \vec r)$. Each \code{Hand} object tracks one hand and is the focal point for recursive generation of hands.

First consider recursive generation of all possible single hands for $E(\vec h,u)$ in \Eq{exactexval}. A simple recursive method in the \code{Hand} object is:

{\small\begin{verbatim}
     float Hand::hitExval(Deck &deck,Dealer &dealer)
     {  float exval=0.,wt;
        for(int i=ACE; i<=TEN; i++)
        {  if(!deck.removeAndGetWeight(i,&wt,dealer)) continue;
           hit(i);
           if(basicHit(deck,dealer))
              exval += wt*hitExval(deck,dealer);
           else
              exval += wt*standExval(deck,dealer);
           unhit(i,deck);
        }
        return exval;
     }
\end{verbatim}}
The loop is over the ten possible cards. \code{deck.removeAndGetWeight()} uses the \code{Deck} object to remove a card from the deck and calculate $p(i)=\code{wt}$; it returns \code{false} if no such cards remain in the deck. The \code{hit()} method adds a card to the hand. If the \code{basicHit()} method says to hit again, the \code{hitExval()}method is called recursively; otherwise the final hand expected value is calculated by \code{standExval()}. Finally the loop card is removed from the hand and restored to the deck (by \code{unhit()}). When the loop is done, the exact expected value is returned (\code{exval}).

A similar method is the core of all blackjack programs for exact expected values. A extension of this approach to recursively generating all possible combinations of split hands, including resplitting, is given by the  \code{exactSplitExval()} method in Appendix B. For splitting a hand with $(s,s)$ against dealer up card $u$, this method is called as follows:

{\small\begin{verbatim}
     deck.remove(u,s,s);
     dealer.setDDAfterSplitOption(option);
     hands[0] = new Hand(s);
     hands[1] = new Hand(s);
     numHands = 2
     exval = hands[0]->exactSplitExval(deck,dealer,hands,numHands,maxHands);
\end{verbatim}}
The first line removes the dealer up card, $u$, and two split cards, $s$, from the deck using a \code{Deck} object. The second line sets the desired double-down-after-split option. Finally, two \code{Hand} objects are created each with a single split card and the calculation is launched in the first hand. \code{maxHands} determines $h$ with $h=2$ for no resplitting or $h>2$ to allow resplitting. The logic of \code{exactSplitExval()} is very similar to \code{hitExval()}. The major difference is that the recursion proceeds through two or more \code{Hand} objects rather than being confined to a single \code{Hand}.

The first calculations considered splitting all possible pairs ((A,A), (2,2), ... (T,T)) {\em vs.} dealer up card 6 and all double down options (ND, DD1, and DD2). The calculations for maximum number of hands $h=1$, 2, or 3, took 0.69 sec, 49 sec, and 8666 sec, respectively (note: the single hand calculation assumes the expected value of splitting ($s$,$s$) is twice the expected value for playing a single hand of ($s$) \cite{Griffin:1979}). All calculations were done using 3~GHz Intel Xeon processors (either a Mac or HP Linux cluster node from 2008). Recursive generation of $h$ hands should  scale roughly as $N^h$ where $N$ is the effective number of possible single hands. The effective $N$ for dealer up card 6 was 73 to 176. Repeating that calculations for dealer up card 9 for for $h=1$ or 2, took 2.75 sec and 4482 secs, for an effective $N=1631$. Calculations for $h=3$ were too slow to attempt.

Calculations for dealer up card 6 (and other low cards 2-6) have lower $N$ and complete faster. Calculations for dealer up card 9 (and other high up cards 7-10, A) have a higher $N$ and take much more time. The reason for lower $N$ {\em vs.\/} low up cards is that basic strategy stops hitting at a low score (12 or 13). Against high up cards basic strategy continues hitting to 17 thus requiring consideration of longer hands. By considering all up cards and extrapolating observed scaling effects, it was estimated that completing splitting calculations for all up cards and all splitting rule options up to $h=4$ would  take $3.4\times 10^{11}$ seconds or 11,000 years of CPU time. The following sections give strategies for faster calculations.

\subsection{Dealer Caching Method}
\label{DealerCachingMethod}

The last step for each generated hand is to calculate its expected value. This step requires enumeration through all possible dealer hands for the current deck composition to find $\vec D(u,\vec r)$. In the process of generating player hands, however, the same deck composition will occur many times. A significant improvement in performance is possible by caching dealer probabilities such that the next time the same composition occurs, $\vec D(u,\vec r)$ can be found in a look-up table. The key task is an addressing mechanism for storing dealer results.

Imagine a table to hold all dealer results for decks with 0 to $j$ cards removed. Because removal order does not matter, the removed cards are first put into decreasing sequence $(x_1,x_2,\dots,x_j)$ with $x_1\ge x_2 \ge x_3 \ge \dots x_j$ and $x_i=0$ to $10$ for card type removed. A value of $x_i=0$ is used to signify no card removed to handle compositions with less than $j$ removed cards. Define $K_j(x_1,x_2,\dots,x_j)$ as the address found by enumerating through all compositions
\[
         K_j(x_1,x_2,\dots,x_j) =  \sum_{i_1=0}^{x_1} \sum_{i_2=0}^{\min(i_1,x_2)} \cdots 
         \sum_{i_j=0}^{\min(i_{j-1},x_j)} 1
\]
until $i_1=x_1$, $i_2=x_2$, $\cdots$, $i_j=x_j$. This address is better defined using a recurrence relation:
\begin{equation}
    K_1(x_1) = 1+x_1 \ {\rm and} \ K_j(x_1,x_2,\dots,x_j) = K_{j-1}(x_2,\dots,x_j) + \sum_{i=0}^{x_1-1}T_{j-1}(i+1)       \label{Kdef}
\end{equation}
where
\[
      T_j(N) = K_j(N-1,N-1,\dots,N-1) 
\]
is the address of the last element in a table required to save $j$ items in which each item can assume $N$ states (0 to $N-1$) and is thus the length of that table.
Evaluation of $K_j(N-1,N-1,\dots,N-1)$ using \Eq{Kdef} gives
\[
      T_j(N) = \sum_{i=0}^{N-1} T_{j-1}(i+1)
\]
which leads to
\begin{equation}
    K_j(x_1,x_2,\dots,x_j) =  1+\sum_{i=1}^{j} T_{i}(x_{j-i+1})     \label{address}
\end{equation}
From the definition of $T_j(N)$ initiated with $T_1(N)=N$:
\begin{equation}
      T_2(N) = {N(N+1)\over 2}, \  T_3(N) = {N(N+1)(N+2)\over 2\cdot3}, 
          \  T_j(N) = \left( \begin{array}{c} N+j-1 \\ j \end{array}\right)       \label{TjN}
\end{equation}

A dealer cache is created by setting $j$ to maximum allowed by available memory. Each time the dealer probabilities are needed, the address is calculated by \Eq{address}. To avoid repeated evaluation of $T_j(N)$, all needed values can be pre-calculated and stored in a $j\times 11$ array. The first time an address is encountered, the dealer probabilities are calculated and stored in the cache. All subsequent times, the probabilities are retrieved and used. The length of the dealer cache is $T_j(11)$. The largest cache used was $j=23$ requiring 2.23 GB of storage (when using standard four-byte floats)

\begin{figure}
\centerline{\includegraphics[width=0.8\textwidth]{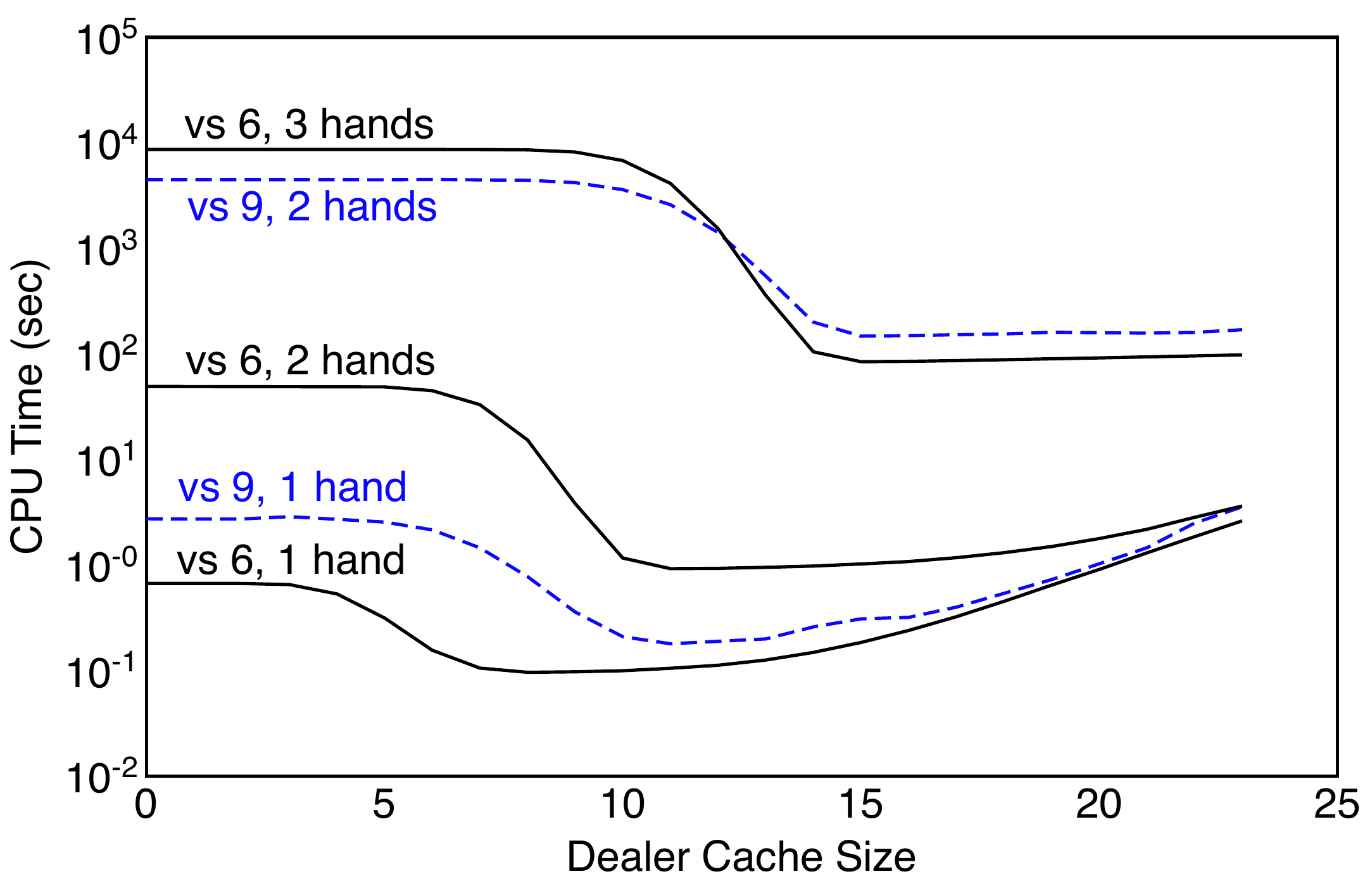}}
\caption{\small The effect of dealer cache size on splitting calculation times for dealer up card 6 (solid lines) or 9 (dashed lines) with total play allowed to 1 hand, 2 hands, or 3 hands. The ``\emph{vs 9, 3 hands}'' were too slow and are not plotted.}
\label{Cache}
\end{figure}

Figure \ref{Cache}\ shows the effect of dealer cache size when evaluating splitting of all pairs with all double down options (ND, DD1, and DD2) for dealer up cards 6 or 9 and various $h$ values (as mentioned above, calculations for high up card of 9 take longer than for low up card 6). Very small caches are no help, but once the cache size reaches the common hand size, the calculation time drops rapidly showing improvements of up to two orders of magnitude. Larger cache sizes are required as $h$ increases. The time increases at very large cache size (\emph{e.g.}, 1 hand results in \Fig{Cache}) if the time needed to allocate and initialize the cache is non-negligible compared to calculation time.

\begin{figure}
\centerline{\includegraphics[width=0.8\textwidth]{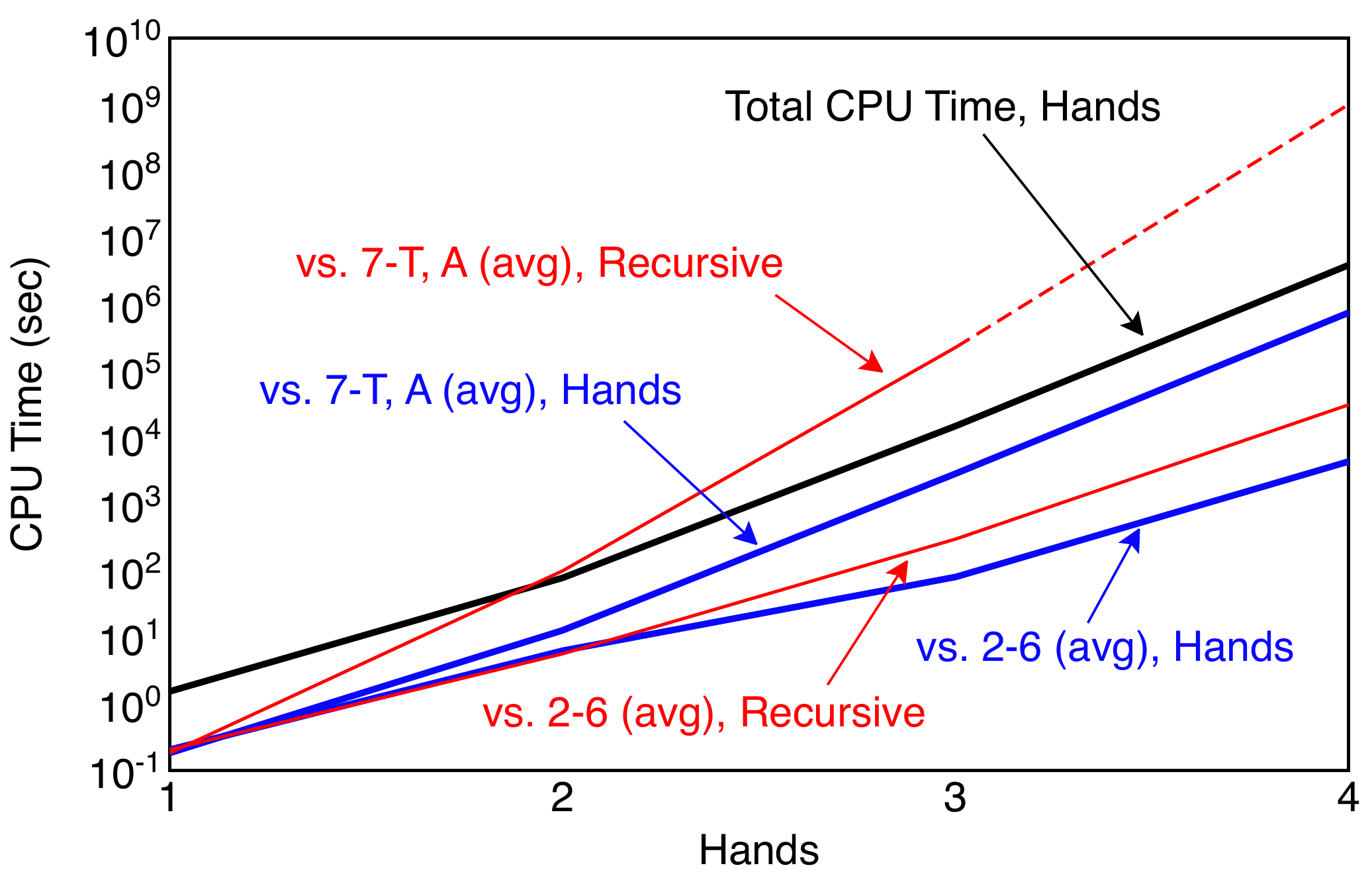}}
\caption{\small Splitting calculation times for various up cards done be either ``Recursive'' methods (thin, red lines) or by pre-enumerated ``Hands'' method (thick, blue lines) as a function of the maximum number of allowed split hands. The solid, black line is the total execution time {\em vs.\/}~all dealer up cards using the pre-enumerated ``Hands'' method.}
\label{Scaling}
\end{figure}

The next calculations reconsidered splitting of all pairs with all double down options (ND, DD1, and DD2) for all possible dealer up cards using optimum cache sizes. Average calculation times for the recursive method with a dealer cache for low ({\em vs.\/}~2-6, Recursive) or high ({\em vs.\/}~7-T and A, Recursive) dealer up cards are plotted as a function of the number of hands, $h$, in \Fig{Scaling} as thin, red lines. The scaling was approximately $N^h$,  with $N$ effectively $N=50$ to 100 for low up cards and over 2200 for higher up cards. Note that $N$ is similar to the first calculations, but the pre-factor for absolute time is significantly reduced. Calculations up to 4 hands with low dealer up cards were completed, but calculations for high dealer up cards were still too slow. The dashed line in \Fig{Scaling}\ extrapolates the time to $h=4$ for high dealer up cards by mimicking the increase in effective $N$ for low up cards. This extrapolation provides and estimate of $5.0\times 10^9$ or 160 years of CPU time for all calculations --- a significant improvement, but still too long.

\subsection{Pre-enumerated Hands Method}
\label{PreenumeratedHandsMethod}

The problem with recursive hand generation is that it is inefficient. The method revisits many identical hands that differ only by order in which the hand receives cards. For example, when splitting 2's {\em vs.\/}~dealer 9, recursive methods generate 21,166 hands, but only 1527 of those hands are unique. The next improvement was to catalog all unique hands and then evaluate splitting expected values by enumerating over only the unique hands.

Cataloging of unique hands was done by recursive generation of all possible single hands. An array of \code{PlayHand} objects was constructed to hold the hands. For each unique hand, the \code{PlayHand} object tracks hand composition, the number of times the hand occurs, the number of times the second card in the hand is another split card, total bet size on all hands, and total bet size on potential split hands. Tracking the second card is needed to allow resplitting calculations. Tracking bet size is needed to support doubling down after splitting options. The dealer-cache-addressing methods were used to efficiently determine whether each generated hand is a new hand or corresponds to a previously generated hand. A hand look-up table of $T_j(11)$ integers, where $j$ is the maximum hand length,{\footnote{The maximum hand length in a split hand played by basic strategy with infinite decks is 14 cards. The hand is (1,1,1,1,1,1,1,1,4,1,1,1,1,1) corresponding to split aces played out as normal hands. The 4 as the ninth card is when hitting soft 18 converts the hand to a hard 12.}}\ was created and filled with -1. Each hand uses  \Eq{address} to find an address from its decreasing card composition. If the table entry at that address is -1, a new \code{PlayHand} object is created and added to the unique hands array. The location of the new hand in that array is stored in a hand look up table. The next time the same hand occurs the look up table will provide its location in the array of unique hands and the new information can be appended to the existing \code{PlayHand} object. Notice that building a unique hands list requires recursive generation of hands. Because splitting calculations revisit the same individual hands many times, a unique hands list provides significant benefit. All other calculations visit each hand only once and therefore a unique hands list provides no computational benefit.

Once the unique hands are collected, the exact splitting expected values are evaluated as above except using the new \code{handExactSplitExval()} method in Appendix C:

{\small\begin{verbatim}
     exval = hands[0]->handExactSplitExval(deck,dealer,hands,
                                           numHands,maxHands,handList);
\end{verbatim}}

The last argument is an array of \code{PlayHand} objects for all possible unique hands for the current splitting rules.
Average calculation times when using a hands list for low ({\em vs.\/}~2-6, Hands) or high ({\em vs.\/}~7-T and A, Hands) dealer up cards are plotted as a function of the number of hands, $h$, in \Fig{Scaling} as thick, blue lines. For low up cards, the calculation times decreased by an order of magnitude for $h>2$ and the effective $N$ was reduced to 12 to 50. For high up cards the times for $h=4$ decreased about three orders of magnitude. The effective $N$ was reduced to 250 and did not increase with $h$. Figure \ref{Scaling}\ also shows the total calculation time to find expected values for all dealer up cards, all splitting pairs, and all splitting options (ND, DD1, and DD2) for a single deck game. Calculations for $h=4$ took $3.9\times 10^6$ sec or 45 days --- five orders of magnitude faster than non-optimized recursive methods. The calculations were done using on a 32-processor cluster by partitioning into different up cards and different splitting pairs. Some calculations took much longer than others. 71.6\%\ of the calculation time was needed for up cards 7, 8, and 9; 43.4\%\ of the time was for splitting of just (2,2) and (3,3) for up cards 7, 8, and 9.

\subsection{Parallelization}

The above algorithm describes a serial calculation of exact expected values for pair splitting that minimizes computer time. The algorithm could possibly be parallelized, but its use of caches for dealer probabilities and unique player hands complicates that task. To achieve good scaling, each processor would need separate caches. By this approach, separate processors would lose the benefit of learning from dealer calculations done in a other processors. In shared-memory parallelization, multiple copies of caches would limit the memory available to each one thus limiting dealer probabilities that could be cached. Multiple processors could share a single cache, but the frequent locks needed to write new results to the cache would reduce scaling performance. The dealer cache, however, applies only to a single dealer up card. The best strategy for parallel code, therefore, is for each processor to do calculations for a single up card. The simplest approach to this parallelization, is simply to run 10 serial calculations with each one considering a single dealer up card.

\section{Results and Discussion}
\label{ResultsandDiscussion}

Tables \ref{exactLow}\  and \ref{exactHigh}\ give exact splitting expected values for all split pairs against all dealer up cards in a single-deck game where the dealer stands on soft 17. Each table cell has four numbers. The first row is when resplitting is not allowed; the second row is when resplitting is allowed (to a maximum of 4 hands). Within each row, the first value is when doubling down after splitting is not allowed; the second value is when doubling down after splitting on any two cards is allowed. Splitting of aces assumes each hand gets a single card. Resplitting of aces means if the single card is another ace, that new pair can be split again, but still only receives a single card. Exact calculations were also completed for dealer hitting soft 17 and for double down after splitting on only 10 and 11. Those results are not included in table, but are included in the total CPU times. Thus total CPU time is the time required to analyze all possible splitting rule variations in a single deck game.

\begin{sidewaystable}
\caption{\small Exact splitting expected values {\em vs.}~dealer up cards Ace through 5 for a single deck game where dealer stands on soft 17. The first row of each cell is when resplitting is not allowed; the second row is when resplitting is allowed to 4 hands. The first column of each cell is when doubling down after splitting is not allowed; the second column is when it is allowed on any two cards. Decimal points in front of each number were omitted to save space.}
\bigskip\label{exactLow}
\begin{center}
{
\begin{tabular}{|c|cc|cc|cc|cc|cc|}
\hline
Hand & \multicolumn{2}{c|}{Ace} & \multicolumn{2}{c|}{2} & \multicolumn{2}{c|}{3}
      & \multicolumn{2}{c|}{4} & \multicolumn{2}{c|}{5} \\
\hline
\multirow{2}{*}{A,A}
               & 223932 & 223932 & 565702 & 565702 & 612855 & 612855 & 668582 & 668582 & 732160 & 732160 \\
                &  261074 & 261074 & 639246 & 639246 & 686841 & 686841 & 742469 & 742469 & 807466 & 807466 \\
\hline
\multirow{2}{*}{2,2}
               & -427834 & -421365 & -128397 & -0393101 & -0701432 & 0290697 & 00318339 & 117502 & 127678 & 273309  \\
               & -442273 & -434935 & -128950 & -0358712 & -0690329 & 0393212 & 00632171 & 131134 & 136489 & 295745   \\
\hline
\multirow{2}{*}{3,3}
               & -471347 & -462243 & -197137	 & -112304 & -127583 & -0275217 & 00933911 & 135241 & 116658 & 276757  \\
                & -481365 & -471167 & -200181 & -107650 & -127644 & -0232815 & 0157778 & 153363 & 127911 & 303241   \\
\hline
\multirow{2}{*}{4,4}
               & -547373 & -535358 & -235941 & -149759 & -123036 & -0158384 & -0133961 & 122532 & 0947167 & 256441  \\
               & -577198 & -563821 & -255764 & -161787 & -134928 & -0177820 & -0174282 & 124641 & 0908038 & 264619 \\
\hline
\multirow{2}{*}{5,5}
               & -648455 & -633159 & -231716 & -156772 & -146072 & -0559217 & -0382401 & 0703683 & 0681172 & 205292 \\
               & -715392 & -698411 & -270361 & -209723 & -179259 & -105188 & -0653697 & 0251227 & 0572212 & 184731 \\
\hline
\multirow{2}{*}{6,6}
               & -637988 & -621394 & -218896 & -121724 & -132663 & -0136772 & -0275570 & 111605 & 0677142 & 237513 \\
               & -658692 & -640104 & -212367 & -105632 & -123591 & 00709888 & -0140747 & 138752 & 0851339 & 271396 \\
\hline
\multirow{2}{*}{7,7}
               & -606933 & -595329 & -162585 & -0728641 & -0789592 & 0295195 & 0162682 & 155107 & 0387914 & 193212 \\
               & -611818 & -598886 & -151898 & -0537095 & -0644500 & 0548212 & 0341069 & 186497 & 0557767 & 223782 \\
\hline
\multirow{2}{*}{8,8}
               & -340317 & -333718 & 0167248 & 0890850 & 0817322 & 161814 & 105595 & 195731 & 188016 & 291005 \\
               & -324889 & -317635 & 0437250 & 122665 & 110432 & 197907 & 133230 & 230970 & 218373 & 330046 \\
\hline
\multirow{2}{*}{9,9}
               & -0721452 & -0705435 & 170069 & 202472 & 170369 & 211550 & 252688 & 299234 & 339220 & 392304 \\
               & -0716589 & -0700780 & 172931 & 208184 & 172686 & 217172 & 258737 & 309099 & 349719 & 407189 \\
\hline
\multirow{2}{*}{T,T}
               & 0985434 & 0985434 & 315676 & 315676 & 364448 & 364448 & 424782 & 424782 & 496722 & 496722 \\
               & -354749 & -354749 & 0473715 & 0473715 & 124252 & 124252 & 224812 & 224812 & 326618 & 326618 \\
\hline
\end{tabular}
}
\end{center}
\label{default}
\end{sidewaystable}%

\begin{sidewaystable}
\caption{\small Exact splitting expected values {\em vs.}~dealer up cards 6 through Ten for a single deck game where dealer stands on soft 17. See caption to Table \ref{exactLow}\ for more details.}
\bigskip\label{exactHigh}
\begin{center}
{
\begin{tabular}{|c|cc|cc|cc|cc|cc|}
\hline
Hand & \multicolumn{2}{c|}{6} & \multicolumn{2}{c|}{7} & \multicolumn{2}{c|}{8}
      & \multicolumn{2}{c|}{9} & \multicolumn{2}{c|}{Ten} \\
\hline
\multirow{2}{*}{A,A}
      & 758276 & 758276 & 540712 & 540712 & 406468 & 406468 & 289770 & 289770 & 194252 & 194252 \\
      & 831965 & 831965 & 627650 & 627650 & 486371 & 486371 & 360781 & 360781 & 259524 & 259524 \\
\hline
\multirow{2}{*}{2,2}
      & 118982 & 272414 & -0544634 & 00475233 & -212597 & -177104 & -383379 & -359454 & -459601 & -451895 \\
      & 126812 & 294721 & -0512111 & 0138648 & -218564 & -179369 & -396823 & -370203 & -475562 & -467058 \\
\hline
\multirow{2}{*}{3,3}
      & 111575 & 274115 & -115327 & -0572504 & -265388 & -226050 & -418153 & -396141 & -504229 & -494166 \\
      & 122171 & 300526 & -108698 & -0448095 & -265841 & -222313 & -426060 & -401782 & -517177 & -506105 \\
\hline
\multirow{2}{*}{4,4}
      & 0834385 & 244333 & -222838 & -158732 & -342754 & -299949 & -493182 & -469024 & -570489 & -559246 \\
      & 0765057 & 248774 & -253997 & -183436 & -369351 & -322316 & -518806 & -492184 & -599527 & -587208 \\
\hline
\multirow{2}{*}{5,5}
      & 0558514 & 200283 & -298829 & -244667 & -448230 & -412012 & -607597 & -584760 & -671747 & -660597 \\
      & 0302934 & 153548 & -352763 & -312307 & -509121 & -481385 & -674606 & -655884 & -736385 & -724109 \\
\hline
\multirow{2}{*}{6,6}
      & -0106513 & 155507 & -269899 & -201370 & -412582 & -367710 & -569709 & -544299 & -655152 & -644208 \\
      & -00389922 & 169340 & -267862 & -192961 & -418612 & -369386 & -584491 & -556572 & -677010 & -664948 \\
\hline
\multirow{2}{*}{7,7}
      & 0530242 & 209924 & -122529 & -0657473 & -423471 & -385238 & -563840 & -545047 & -625405 & -612578 \\
      & 0727575 & 243348 & -110318 & -0512847 & -422274 & -380591 & -568856 & -548340 & -636688 & -622459 \\
\hline
\multirow{2}{*}{8,8}
      & 233716 & 340370 & 202329 & 250712 & -100087 & -0735091 & -429934 & -406325 & -462307 & -452319 \\
      & 269513 & 385294 & 251584 & 303664 & -0869704 & -59524.5 & -427063 & -401144 & -458498 & -447501 \\
\hline
\multirow{2}{*}{9,9}
      & 359377 & 413268 & 340574 & 362756 & 179367 & 195409 & -112424 & -102933 & -265721 & -260551 \\
      & 365950 & 424280 & 334969 & 358600 & 190276 & 207491 & -108836 & -0990434 & -277966 & -272453 \\
\hline
\multirow{2}{*}{T,T}
      & 525105 & 525105 & 482229 & 482229 & 353853 & 353853 & 183754 & 183754 & 0600557 & 0600557 \\
      & 363571 & 363571 & 251783 & 251783 & 0117653 & 0117653 & -255830 & -255830 & -316452 & -316452 \\
\hline
\end{tabular}
}
\end{center}
\label{default}
\end{sidewaystable}%

Prior to these exact calculations, the best splitting calculations used the approximate methods developed by Griffin \cite{Griffin:1979}. If $\vec h_i$ for $i=1$ to $N$ is the collection off all possible unique hands for one splitting situation, the exact expected value used in the hands-lines method for splitting a pair of $s$ cards when resplitting is not allowed can be written explicitly as
\begin{equation}
        E_s(\vec h(s,s),u) =  \sum_{i=1}^{N}  \sum_{j=1}^{N} w(\vec h_i)w(\vec h_j|\vec h_i)\biggl[b(\vec h_i)S(\vec h_i,u,\vec h_j)+b(\vec h_j)S(\vec h_j,u,\vec h_i)\biggr]      \label{exNonRS}
\end{equation}
Here $w(\vec h_i)$ is the weight for hand $i$ defined as the probability of hand $i$ times the number of times that unique hand occurs, $w(\vec h_j|\vec h_i)$ is the analogous weight for hand $j$ given that hand $i$ has been removed from the deck, $b(\vec h)$ is the average bet size for a hand, and $S(\vec h_i,u,\vec h_j)$ is the expected value for standing on hand $i$ against dealer up card $u$ given that both hand $i$ and $j$ have been removed from the deck. Note that $b(\vec h)=1$ when doubling down after splitting is not allowed, but $b(\vec h)> 1$ when it is allowed and one or more occurrences of the hand should be doubled. Griffin's approximate result \cite{Griffin:1979}\ can be expressed as
\begin{equation}
        E_s(\vec h(s,s),u) \approx 2 E(\vec h(s),u,s) = 2\sum_{i=1}^{N} w(\vec h_i)b(\vec h_i)S(\vec h_i,u,s)        \label{appNonRS}
\end{equation}
where $E(\vec h(s),u,s)$ is the expected value of playing out a hand with a single card $s$ against a dealer up card given that the second $s$ card has also been removed from the deck. The exact and approximate methods agree within $\pm 0.000003$ for all splitting calculations. Hawkins \cite{Hawkins:2000}\ compared one exact non-resplitting calculation to Griffin's approximate formula and concluded the similarity implies the approximate formula is exact. Comparison of \Eq{exNonRS}\ to \Eq{appNonRS}, however, shows they are only identical if
\[ 
          S(\vec h_i,u,s) =  {1\over 2} \sum_{j=1}^{N} w(\vec h_j|\vec h_i)\biggl[S(\vec h_i,u,\vec h_j)
                        +{b(\vec h_j)\over b(\vec h_i)}S(\vec h_j,u,\vec h_i)\biggr]
\]
Because averaging over all hands $\vec h_j$ is expected to average out variations in $S(\vec h_i,u,\vec h_j)$, the equation is likely to be accurate. Calculations show it is extremely accurate, but the approximate approach is not an exact result.

Calculations when resplitting is allowed are much more complicated. The state-of-the-art, prior to these exact results, is again due to Griffin \cite{Griffin:1979}. The approximate equation is
\[
     E_s(\vec h(s,s),u) \approx \sum_{i=2}^{h} i P(i)E(i)
\]
where $P(i)$ is the probability of playing exactly $i$ hands, $h$ is the maximum number of hands, and $E(i)$ is a single hand expected value calculated as follows:
\begin{enumerate}
\item Calculate the expected value for a single hand starting with an $s$ card for a deck with the dealer up card and $i$ of the $s$ cards removed from the deck.
\item The player hand cannot draw an $s$ card as the second card in the hand.
\item The dealer expected values are found conditional that the $i-1$ additional player hole cards are not $s$ cards.
\end{enumerate}
The third requirement means dealer expected values calculated for this analysis differ from all other calculations and thus results stored in the dealer cache cannot be used. The probability of card $k$ for dealer play when finding $E(i)$ for splitting $s$ cards is
\[ \renewcommand{\arraystretch}{1.5}
    p(k,i,s) = \left\{ \begin{array}{ll}
        {\ds n_k\over \ds n_{deck}-i+1} & {\rm for}\ k=s  \\
       {\ds n_k\over \ds n_{deck}-i+1}\left({\ds n_{deck}-n_{s}-i+1\over \ds n_{deck}-n_{s}}\right)
                & {\rm for}\ k\ne s
         \end{array} \right.
\]
For calculations to a maximum of four hands ($h=4$), $E(4)$ allows the hand to draw an $s$ for its second card, the dealer expected values are unaffected by player hole cards, and $P(4)$ is set to $P(4)=1-P(2)-P(3)$. 

This approximate analysis was compared to exact results. For splitting aces, and resplitting 2 through 9, the approximate equation is very accurate. Nearly all results agree within $\pm 0.0010$. The few exceptions are splitting 4's {\em vs.\/}~5 and 6 and splitting 5's {\em vs.\/}~4 and 6. The largest error was $\pm 0.0016$. The results for resplitting 10's were not as good. The errors ranged up to $\pm 0.1000$. The problem is that $P(4)$ is much larger and the calculation of $E(4)$ is less accurate.

A new approximate analysis was developed that significantly improves on Griffin's approach when resplitting 10's and also improves the accuracy for all other calculations. When $h=4$, it takes 4 cards to resolve the hand play. Let $s$ be a split card and $n$ be a non-split card. The 16 possible orders for $n$ and $s$ cards in 4 cards are classified as in Table \ref{cardOrder}.
For example, the first row is all combinations where the first two cards are non-split cards. These two non-split cards mean exactly two hands will be played. The approximate expected value is $2E(2)$ and the probability of those four sequences  in $P(2)$. The remaining lines partition 3- and 4-hand situations to refine approximate calculations. For example, the second row has order $nsnn$. The first $n$ means the first hand is played when only two $s$ cards have been removed from the deck and thus has approximate expected value $E(2)$. The next $s$ means the second hand is split; the following two $n$'s mean those two hands are played resulting in three final hands. Since the last two hands are played with three $s$ card removed, their approximate expected values are $E(3)$ each. The expected value of the second row is this $E(2)+2E(3)$. The probability is denoted as $P(3/1)$. Similar logic applies to all remaining rows. Finally, a revised approximate expected value for splitting when resplitting is allowed is
\begin{eqnarray}
      E_s(\vec h(s,s),u) & \approx  & \Bigl[2P(2)+P(3/1)+P(4/1)+P(4/2)\Bigr]E(2) \nonumber\\
 && \ \mbox{}
     + \Bigl[3P(3) - P(3/1)+P(4/1)+2P(4/3)+P(4/4)\Bigr]E(3) \nonumber\\
 && \ \mbox{}
     + \Bigl[4P(4) - 2P(4/1) - P(4/2)-2P(4/3)-P(4/4)\Bigr]E(4)     \label{approxSplit}
\end{eqnarray}
Notice that this new approximation still depends only on $E(2)$, $E(3)$, and $E(4)$ and thus is no more work than Griffin's method. It simply revises the probabilities associated with each expected value. These probabilities can be easily calculated by methods in Griffin (see Appendix D) and it is noted that $P(3) = P(3/1)+P(3/2)$, and $P(4)=P(4/1)+P(4/2)+P(4/3)+P(4/4)+P(4/5)$.
This revised approximate analysis improves agreement for resplitting 10's with exact results by more than an order of magnitude. The largest errors are reduced from $\pm 0.100$ to $\pm 0.006$ with many errors under $\pm 0.004$. 

\begin{table}
\caption{\small All sequences four cards of type $n$ (for non-split card) and $s$ (for split card) and the approximate expected value and probability appropriate for that sequence.}
\bigskip\label{cardOrder}
\begin{center}
\begin{tabular}{lll}
Card Order & Approximate Expected Value & Probability \\ \hline
$nnnn$, $nnns$, $nnsn$, $nnss$ & $2E(2)$ & $P(2)$ \\
$nsnn$ & $E(2) +2E(3)$ & $P(3/1)$   \\
$snnn$ & $3E(3)$ & $P(3/2)$   \\
$nsns$ & $E(2)+E(3)+2E(4)$ & $P(4/1)$  \\
$nssn$, $nsss$ & $E(2)+3E(4)$ & $P(4/2)$ \\
$snns$ & $2E(3)+2E(4)$ & $P(4/3)$  \\
$snsn$, $snss$ & $E(3)+3E(4)$ & $P(4/4)$ \\
$ssnn$, $ssns$, $sssn$, $ssss$ & $4E(4)$ & $P(4/5)$
\end{tabular}
\end{center}
\end{table}

Although many books and articles give splitting strategy, very few references provide splitting expected values. The first are tables in Thorp's {\em Beat the Dealer} \cite{Thorp:1962}. His tables give expected values (to three significant digits) for all pairs against all up cards for a single deck game where resplitting is not allowed, doubling down after splitting is allowed on any two cards, and the dealer stands on soft 17. Epstein \cite{Epstein:1977}\ gives essentially the same expected values (but to four significant figures). Although Epstein claims to have analyzed resplitting allowed, his tables are much closer to results when resplitting is not allowed. The mean errors in these tables compared to the exact results for no replitting are $\pm 0.022$ with some errors $\pm 0.07$. Neither  reference gives sufficient information to explain the discrepancy. Epstein states an approximate formula analogous to \Eq{appNonRS}, but since that equation is accurate, he must have used it incorrectly. As noted by Griffin \cite{Griffin:1979}, the single hand expected value for splitting calculations should be found with the second card removed from the deck. To check if Thorp and Epstein omitted this refinement (since their work was prior to Griffin \cite{Griffin:1979}), new calculations were run without removing the second card. These new results cut the Thorp and Epstein mean error in half to $\pm 0.011$ and maximum error to $\pm 0.027$. In summary, these historic table are not particularly accurate, probably used the wrong approximate formula, and only provide results for one set of splitting rules.

Manson {\em et al.\/}~\cite{Manson:1975}, did calculations for a four deck game with resplitting allowed, doubling down allowed on any two cards, and dealer stands on soft 17. They estimated that exact calculations would take 100 times longer than their other calculations (an estimate that is off by about 9 orders of magnitude), and therefore did all splitting calculations by Monte Carlo simulations. Their table only gives splitting expected values for hands in which splitting is favorable. Because exact results for multiple decks have not be done yet, these results were compared instead to the new approximate analysis for resplitting in four  decks to four hands. The mean absolute error of all their available calculations is $\pm 0.019$; the maximum error is $\pm 0.04$. These errors are similar to independent Monte Carlo simulations run during this work and probably representative of expected errors for a simulation approach.

Hawkins \cite{Hawkins:2000}\ gives results for a six deck game with no doubling down after splitting and dealer hits soft 17. His analysis used the Griffin \cite{Griffin:1979}\ approximate methods and gives expected values for either resplitting allowed or resplitting not allowed, depending on which was more favorable. When the table gives non-resplitting results, the results are usually identical to calculations by \Eq{appNonRS}. The two exceptions are {\em vs.\/}~dealer up cards ace and ten and for splitting (3,3) {\em vs.\/}~dealer up card 8. The inaccuracies for dealer up card ace and ten suggest Hawkins did not follow Griffin's method to account for the effect on player card probabilities of the knowledge that the dealer does not have blackjack \cite{Griffin:1979}. Hawkins results for  splitting (3,3) {\em vs.\/}~8 were off by exactly a factor of 2 suggesting it was a misprint ({\em i.e.}, omitted the 2 in \Eq{appNonRS}). Hawkins resplitting results were compared to the new approximate analysis for resplitting; they agree well with a mean absolute error of $\pm 0.0005$.

As indicated in Hawkin's table \cite{Hawkins:2000}, and in these new results, if splitting once is favorable, then resplitting is more favorable. Conversely, if the first split is unfavorable, resplitting is more unfavorable. For example, the expected values for splitting tens to a maximum of one hand ({\em i.e.}, standing on the twenty), two hands, three hands, or four hands, {\em vs.\/}~dealer up card of 6 are 0.697403, 0.525105, 0.426506, and 0.363571. In basic strategy, it is never favorable to split tens and thus the decreasing expected values are not relevant. In card-counting strategies, however, one can calculate a cutoff count for splitting tens \cite{Griffin:1979}. When the count exceeds that cutoff value, splitting tens becomes favorable. But, the cutoff values to continue splitting to more hands will be higher than the cutoff value for the first split. Furthermore, because additional tens that allow resplitting lower standard counts, it is highly unlikely for the count to ever be high enough to recommend splitting tens more than once or twice. A common story in card-counting books is about using a count to split and resplit tens \cite{Uston:1981,Mezrich:2002}. In these stories, the first split may have been correct, but it is likely each resplit is giving advantage back to the house. The stories are meant to convey card-counting prowess, but rather show a misunderstanding of the expected values for resplitting. An extreme example in Uston \cite{Uston:1981}\ describes splitting tens up to 11 hands resulting in a \$5000 loss. This outcome was attributed to bad luck despite ``correct plays.'' More likely, the correct play was to split only once or twice.

Finally, by combining exact splitting results with easily-calculated exact results for standing, hitting, and doubling down  \cite{Nairn:BJGit,Nairn:BJWeb}, one can calculate the exact expected value for basic-strategy blackjack including the influence of all splitting rule variations. Such calculations for a single-deck game when dealer stands on soft 17 are summarized in Table \ref{gameEx}. The left half of the table is for games that allow doubling down on any hand with the two columns for whether or not doubling down is also allowed after splitting. The first row is a baseline calculation when resplitting is not allowed. The next two rows give the \emph{change} in expected value by allowing resplitting except for aces and by allowing resplitting including aces. For comparison, the last row gives the \emph{change} in expected value by prohibiting splitting. The same calculations for games that restrict doubling down to 10 and 11 are in the right half of the table. The player benefits of allowing resplitting are small and range from 0.0170\%\ to 0.1889\%. Similarly, the penalty to the player of eliminating the  splitting option is only 0.3566\%\ to 0.3595\%. This benefit of resplitting and penalty of eliminating splitting are small because the total number hands worth splitting is small. Although 11.8\%\ hands could be split, most are hands that should not be split ({\em e.g.}, (10,10)). Under favorable rules, only 2.5\%\ of hands provide beneficial splitting opportunities. The game calculations were repeated when dealer hits soft 17. All expected values decrease by 0.19\%\ to 0.20\%\ (\emph{i.e.}, hitting soft 17 is less  favorable to the player). Because all expected values change the same amount, the effects of different splitting rules are essentially unchanged.

\begin{table}
\caption{\small Effect of splitting rules on total expected values for single-deck blackjack (in percent) when dealer stands on soft 17. The two halves are for doubling down on any two hands or doubling down on just 10 \&\ 11 with two columns in each section for whether or not doubling down is also allowed after splitting. The first row gives baseline game when resplitting is not allowed. The next three rows give the \emph{change} in expected value by allowing resplitting (but not aces), by allowing resplitting (including aces), and by eliminating the splitting option. Entries in {\bf bold} are when the game is favorable to the player.}
\label{gameEx}
\begin{center}
{
\begin{tabular}{|l||r|r||r|r||}
\hline
 \multirow{2}{*}{Options}& \multicolumn{2}{|c||}{DD Any Hand} &  \multicolumn{2}{c||}{DD 10 \& 11 Only} \\
\cline{2-5}
  & \multicolumn{1}{c|}{ND} & \multicolumn{1}{c||}{DD1} & \multicolumn{1}{c|}{ND} & \multicolumn{1}{c||}{DD2} \\
\hline
Base --- No Resplit & -0.0283 & {\bf 0.1007}  & -0.2964  & -0.1904  \\
\hline
$\Delta$ Resplit (-A's) & +0.0170  & {\bf +0.1577}  & +0.0170  & +0.1327  \\
$\Delta$ Resplit (+A's) & {\bf +0.0482}  & {\bf +0.1889}  & +0.0482 & +0.1638  \\
$\Delta$ Eliminate Splitting & -0.3566 & & -0.3595  &  	\\
\hline
\end{tabular}
}
\end{center}
\end{table}%

Single-deck blackjack is a remarkably fair game with the non-resplitting game ranging from 0.2964\%\ house advantage to a 0.1007\%\ player advantage (the table entries in bold are games that provide a player advantage). These calculation were based on exact computer calculations (tabulated to six-digit precision) and therefore very accurate. Repeating game calculations using the new approximate splitting in Eq.~{\ref{approxSplit}\ gives essentially the same results with maximum expected value difference of only 0.00039\%. Because the approximate splitting analysis is extremely accurate, it can be used to calculate expected values in multi-deck games. All expected values decrease about 0.35\%, 0.51\%, and 0.56\% for two-, four-, and six-deck games. That largest incremental decrease is between one- and two-deck games. Thereafter, the effect of more decks has diminishing penalty and approaches a limit of about 0.60\% decrease for an infinite number of decks.

Note that the first significant digit in some expected values is the second digit after the decimal point for percentages (or in the fourth digit for absolute expected values). For these result to be accurate in that digit, tabulated expected values need more digits of precision. Figure~\ref{EVDigits}\ plots game expected value (game rules indicated on the figure) as a function of digits of precisions in tabulated expected value tables for standing, hitting, doubling down, and splitting. This plot shows that tables need at least five digits of precision. Prior published data to three \cite{Thorp:1962} or four significant digits \cite{Epstein:1977}\ might lack precision in full-game expected values. Tables \ref{exactLow} and \ref{exactHigh}\ provide six digits of precision, which is enough precision for accurate game expected value calculations.

\begin{figure}
\centerline{\includegraphics[width=0.7\textwidth]{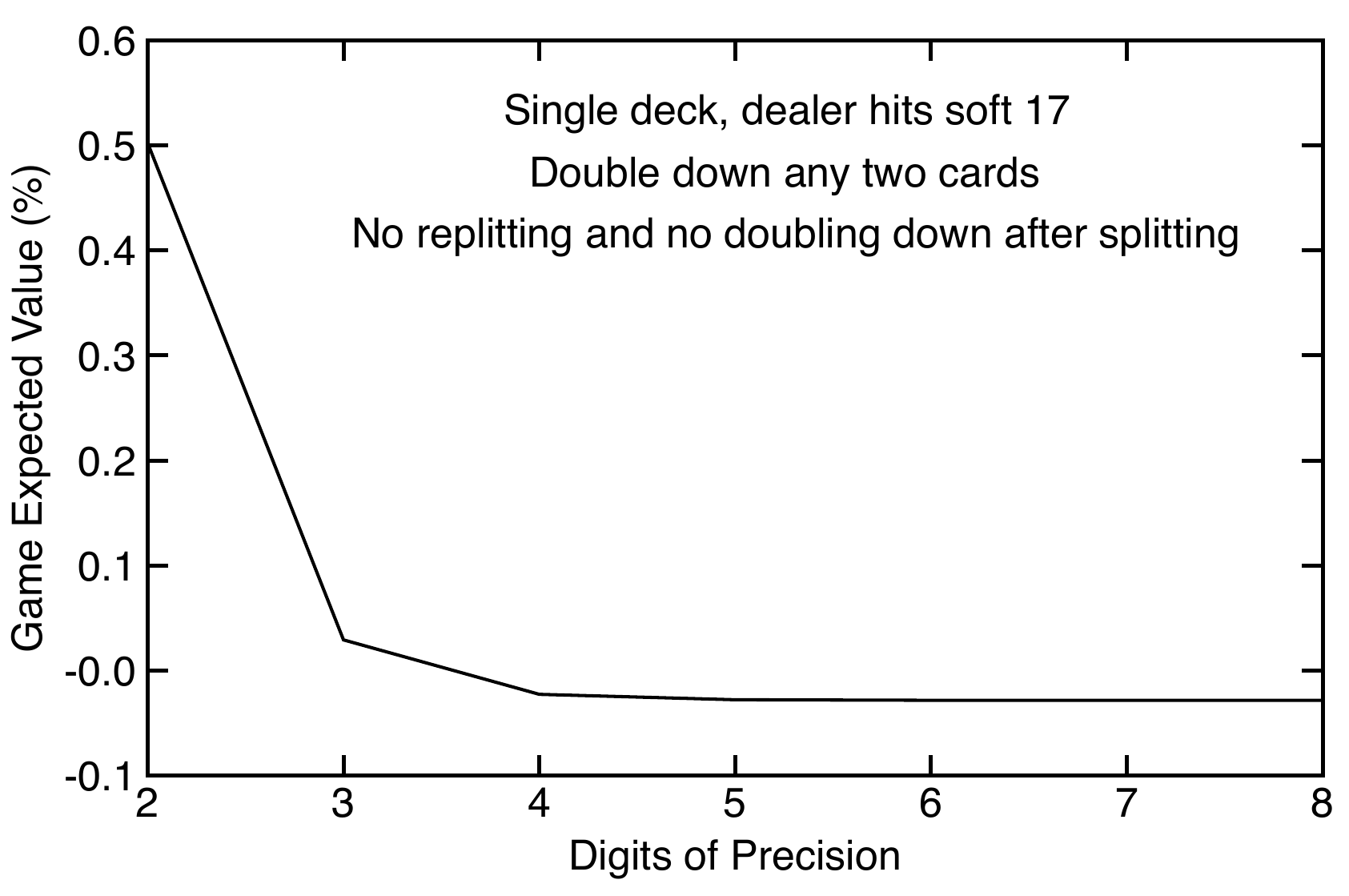}}
\caption{\small Calculation of total game expected value based on expected value tables reported to different number of digits of precision (in the absolute expected values). The analyzed game rules are indicated on the figure.}
\label{EVDigits}
\end{figure}

\section{Conclusions}
\label{Conclusions}

The calculation of exact splitting expected values in a one deck game when resplitting is not allowed or when resplitting is allowed to a maximum of four hands were both solved by computer methods. The calculations when resplitting is not allowed can be done reasonably fast. The results showed that prior approximate calculations  are extremely accurate. The calculations when resplitting is allowed are time consuming, but the methods here make them possible given sufficient computer power. The new exact results showed that prior approximate methods were inaccurate for resplitting 10's. A new approximate solution was developed. It can be done rapidly and is accurate to $\pm .001$ for all resplitting calculations except for resplitting 10's. The approximate analysis for replitting 10's is accurate to $\pm 0.006$.

The overall effect of splitting on blackjack expected values is rather small. These new calculations do not significantly alter previously calculated expected values and do not recommend any changes to previously published strategies for splitting. Solving for exact splitting expected values, however, was an interesting computer challenge. It provides an example of how elegant coding, here represented by a simple recursive algorithm, may be inefficient. Whenever computer recursion is too repetitive, significant improvements may be possible by revising enumeration methods.

Although exact calculations for splitting are lengthy, if all rule variations are included, they only need to be done once. A complete table requires twelve calculations for each split pair \emph{vs.}~each dealer up card --- three double down options (ND, DD1, and DD2) when the dealer hits soft 17, the same three options when the dealer stands on soft 17, and these six calculations have to be done for both resplitting not allowed and resplitting allowed. The tables in this paper only include four of these twelve results for a one-deck game. They include results are for ND and DD1, for resplitting allowed or not allowed, and only for dealer stands on soft 17. The remaining exact calculations for one-deck games are in the software project \cite{Nairn:BJGit} and posted on the web \cite{Nairn:BJWeb}. The exact calculations for multideck games have not been done and would require somewhat more computer time. Multideck results for all twelve calculations, however, were run using the new approximate methods; these results are also in the software project \cite{Nairn:BJGit} and posted on the web \cite{Nairn:BJWeb}.

\appendix
\section*{Appendices}

\subsection*{A. Basic Strategy}

The single-deck basic strategy for hitting and doubling down used in these calculations was

\begin{enumerate}
\item Hard Hitting: Hit 12 or less {\em vs.\/}~dealer up card 2 or 3 (except hit (10,3) {\em vs.\/}~2 if dealer stands on soft 17 and stand on (8,4), (7,5), and (6,6) {\em vs.\/}~3), hit 11 or less {\em vs.\/}~4 to 6 (except hit (10,2) {\em vs.\/}~4 and hit (10,2) {\em vs.\/}~6 if dealer stands on soft 17), and hit 16 or less {\em vs.\/}~7 to 10 or Ace (except stand on (7,7) {\em vs.\/}~10 and stand on 16 with 3 or more cards {\em vs.\/}~10).
\item Soft Hitting: Hit soft 17 or less {\em vs.\/}~dealer up card 2 or 8, hit soft 18 or less {\em vs.\/}~9, 10 and Ace (except stand soft 18 {\em vs.\/}~Ace if dealer stands on soft 17).
\item Hard Doubling Down: Double 9 to 11 {\em vs.\/}~dealer up card 2 to 4, double 8 to 11 {\em vs.\/}~5 and 6 (except hit (6,2) {\em vs.\/}~5 if dealer stands on soft 17 and hit (6,2) {\em vs.\/}~6), double 10 and 11 {\em vs.\/}~7 to 9, and double 11 {\em vs.\/}~10 and Ace.
\item Soft Doubling Down: Double (A,6) {\em vs.\/}~2, double (A,6) and (A,7) {\em vs.\/}~3, double (A,2) to (A,7) {\em vs.\/}~4 and 5, and double (A,2) to (A,8) {\em vs.\/}~6.
\end{enumerate}
The basic strategy for splitting is not needed in the calculations but can be determined from the results.

\subsection*{B. Recursive Hand Generation}

Exact expected values by enumerating through all possible split hands is accomplished by the following algorithm (see software project for full code \cite{Nairn:BJGit}):

{\small\begin{verbatim}
  float Hand::exactSplitExval(Deck &deck,Dealer &dealer,Hand **hands,
                               int &numHands,int maxSplitHands)
  {  float exval = 0., wt, totalVal;
     bool newHand = false;
	
     for(int i=TEN; i>=ACE; i--)
     {  if(!deck.removeAndGetWeight(i,&wt,dealer)) continue;
	
        // add new hand or new card
        if(i == firstCard && cards == 1 && numHands < maxSplitHands)
        {  hands[numHands++] = new Hand(firstCard);
           newHand = true;
        }
        else
           hit(i);

        // hit again, continue to next hand, or add to probabilities
        if(basicSplitHit(deck,dealer))
           exval += wt*exactSplitExval(deck,dealer,hands,
                                        numHands,maxSplitHands);
        else if(this != hands[numHands-1])
           exval += wt*nextHand->exactSplitExval(deck,dealer,hands,
                                       numHands,maxSplitHands);
        else
        {  totalVal=0.;
           for(int j=0; j<numHands; j++)
              totalVal += hands[j]->getExpectedWin(deck,dealer);
            exval += wt*totalVal;
        }
      	
        // delete new hand or unhit and then return card to the deck
        if(newHand)
        {  delete hands[numHands--];
           newHand=false
        }
        else
           unhit(i);
        deck.restore(i);
     }
     return exval;
  }
\end{verbatim}}

The loop is over the ten possible cards. \code{deck.removeAndGetWeight()} uses the \code{Deck} object to remove a card from the deck and calculate $p(i)=\code{wt}$; it returns \code{false} if no such card remains in the deck. If the current card is another split card (and is the second card in the hand), a new hand is created providing the total number of hands in less than $h$ (in \code{maxSplitHands)}, otherwise the \code{hit()} method adds the card to the hand. If the \code{basicSplitHit()} method says to hit again, the method is called recursively; the \code{basicSplitHit()} method also checks if the hand should be doubled down. If the hand is not hit and the current hand is not the last hand, the \code{exactSplitExval()} method is called in the next hand object; otherwise for the last hand, the results for all hands, accounting for doubled down hands, are summed and added with proper weighting to the expected value. If a new hand was created, it is deleted; otherwise the card is removed from the hand (by \code{unhit()}). Finally, the loop card is restored to the deck. When the loop is done, the exact expected value is returned (\code{exval}).

\subsection*{C. Pre-Enumerature Hands Method}

Exact expected values using a previously calculated list of all possible hands (in \code{handList}) is accomplished by the following algorithm (see software project for full code \cite{Nairn:BJGit}):

{\small\begin{verbatim}
  float Hand::handExactSplitExval(Deck &deck,Dealer &dealer,Hand **hands,
                  int &numHands,int maxSplitHands,handset &handList)
  {
     float exval = 0., wt, totalVal;

     // check for resplitting
     if(numHands<maxSplitHands)
     {  if(deck.removeAndGetWeight(firstCard,&wt,dealer))
        {  hands[numHands] = new Hand(firstCard);
           hands[numHands-1]->setNextHand(hands[numHands++]);
           exval += wt*handExactSplitExval(deck,dealer,hands,
                          numHands,maxSplitHands,handList);
           delete hands[--numHands];
           deck.restore(firstCard);
        }
     }

     // hand loop	
     for(int i=0; i<handList.size(); i++)
     {  // get non-splitable fraction or entire hand
        if(numHands<maxSplitHands && handList[i]->isSplitable())
        {  if(!handList[i]->removeAndGetNonsplitWeight(deck,dealer,&wt))
              continue;
           handList[i]->fillNonsplitHand(this);
        }
        else
        {  if(!handList[i]->removeAndGetWeight(deck,dealer,&wt))
              continue;
           handList[i]->fillHand(this);
        }

        // add to probabilities or continue to next hand
        if(this==hands[numHands-1])
        {  totalVal = 0.;
           for(int j=0; j<numHands; j++)
              totalVal += hands[j]->getExpectedSplitWin(deck,dealer);
           exval += wt*totalVal;
        }
        else
        {  exval += wt*nextHand->handExactSplitExval(deck,dealer,hands,
                       numHands,maxSplitHands,handList);
        }
		
        // remove all cards from hand and restore to the deck
        handList[i]->removeHand(this,deck);
     }
     return exval;
  }
\end{verbatim}}

This method is similar to the recursive \code{exactSplitExval()} in Appendix B; the important differences are as follows. The \code{handExactSplitExval()} method is called with an array of \code{PlayHand} objects in \code{handList}; these objects have all possible hands that can be played in the current splitting calculation. Because the hands list is for playable hands, the method has to begin with a separate check for resplitting opportunities. If it is possible to resplit, a new hand is created, \code{handExactSplitExval()} is called recursively, and then the new hand is deleted.

The main loop is over all possible hands rather than over possible cards. The first steps are to remove all cards in the next playable hand, get the weight for that hand, and populate the current \code{Hand} object with those cards. The key to these steps is to account for playable hands that may include one or more hands that could be resplit; {\em i.e.}, one or more of the hands described by the \code{PlayHand} object is a hand where the second card is another split card. When the current number of hands is less the the maximum and the current \code{PlayHand} object includes hands than can repslit, the weighting must be adjusted to include just the fraction of the hand that cannot be resplit; otherwise the entire \code{PlayHand} content is used. The \code{removeAndGetNonsplitWeight()} or \code{removeAndGetWeight()} functions remove all cards in the \code{PlayHand} object from the deck and adjust the weight according to the frequency of that hand; the methods return false if the hand is not possible from the current remaining cards. The \code{fillNonsplitHand()} and \code{fillHand()} methods add the cards to the current \code{Hand} object and also calculate the average bet per hand to adjust for frequency of doubled down hands in the current \code{PlayHand} object.

Next, if the current hand is the last hand, the total expected win for all hands is summed and added to total expected value. The \code{getExpectedSplitWin()} function adjusts the win rate to average bet size per hand to allow calculations with doubling down after splitting. If the current hand is not the last hand, control is passed to the next player hand. Finally, all cards in the current \code{PlayHand} object are removed from the current \code{Hand} object and restored to the deck. When the loop is done, the exact expected value is returned (\code{exval}). This revised enumeration approach was verified as correct by comparison to the direct recursive method in Appendix B when that method was not too slow to be used.

\subsection*{D. Probabilities for Approximate Splitting Calculations}

Define $p_j(s)$ as the probability that the first split card is drawn in the $j^{th}$ position and $p_j(s|s)$ as the probability that the second split card is drawn in the $j^{th}$ position. They are calculated from player card probabilities in \Eq{playerpi}:
\begin{eqnarray*}
      p_1(s) & =  p(s) \\
      p_2(s) & =  p(s)\ {\rm given}\ n_{deck}\to n_{deck}-1 \\
      p_2(s|s) & =  p(s)\ {\rm given}\ n_s \to n_s-1,\ n_{deck}\to n_{deck}-1  \\
      p_3(s) & =  p(s)\ {\rm given}\ n_{deck}\to n_{deck}-2 \\
      p_3(s|s) & =  p(s)\ {\rm given}\ n_s \to n_s-1,\ n_{deck}\to n_{deck}-2  \\
      p_4(s) & =  p(s)\ {\rm given}\ n_{deck}\to n_{deck}-3 \\
      p_4(s|s) & =  p(s)\ {\rm given}\ n_s \to n_s-1,\ n_{deck}\to n_{deck}-3
\end{eqnarray*}
The required probabilities for approximate splitting calculations become:
\begin{eqnarray*}
     P(2) & =  \Bigl[1-p_1(s)\Bigr] \Bigl[1-p_2(s)\Bigr]  \\
     P(3/1) & =  \Bigl[1-p_1(s)\Bigr] p_2(s) \Bigl[1-p_3(s|s)\Bigr] \Bigl[1-p_4(s|s)\Bigr]   \\
     P(3/2) & =  p_1(s) \Bigl[1-p_2(s|s )\Bigr] \Bigl[1-p_3(s|s)\Bigr] \Bigl[1-p_4(s|s)\Bigr]   \\
     P(4/1) & = \Bigl[1-p_1(s)\Bigr] p_2(s) \Bigl[1-p_3(s|s)\Bigr] p_4(s|s)  \\
     P(4/2) & =  \Bigl[1-p_1(s)\Bigr] p_2(s) p_3(s|s) \\
     P(4/3) & =  p_1(s) \Bigl[1-p_2(s|s )\Bigr] \Bigl[1-p_3(s|s)\Bigr] p_4(s|s)  \\
     P(4/4) & =  p_1(s) \Bigl[1-p_2(s|s )\Bigr] \Bigl[1-p_3(s|s)\Bigr] \\
     P(4/5) & =  p_1(s) p_2(s|s)  
\end{eqnarray*}

\bibliographystyle{unsrt}
\bibliography{Splitting_refs} 

\end{document}